\documentclass[pra,showpacs,psfig]{revtex4}

\usepackage{graphicx}
\begin{document}
\title{Local and global statistical distances are equivalent on pure states}
\author{Scott M. Cohen}
\thanks{email: cohensm@duq.edu}
\affiliation{
Department of Physics, Duquesne University, Pittsburgh, PA  15282 \\
\centerline{and} \\
Department of Physics, Carnegie-Mellon University, Pittsburgh, PA 15213
}
\date{\today}
\begin{abstract}

\noindent The statistical distance between pure quantum states is obtained by finding a measurement that is optimal in a sense defined by Wootters. As such, one may expect that the statistical distance will turn out to be different if the set of possible measurements is restricted in some way. It nonetheless turns out that if the restriction is to local operations and classical communication (LOCC) on any multipartite system, then the statistical distance is the same as it is without restriction, being equal to the angle between the states in Hilbert space. 

\end{abstract}

\maketitle

\section{Introduction}

To what extent do two states of a given quantum system differ from each other? The answer to this question is central to our understanding of quantum information, and of quantum physics in general. In particular this difference, if defined appropriately, can be thought of as a ``distance" between the two states --- call it $d(\psi^{(1)},\psi^{(2)})$ when the states are $|\psi^{(1)}\rangle$ and $|\psi^{(2)}\rangle$ --- and this distance can be used to introduce notions of geometry in considerations of the (Hilbert) space in which the states reside \cite{BraunsteinCaves,ZyczGeometry}. However, it is not immediately obvious how to define such a distance, and there could be a variety of acceptable definitions that one might use. A reasonable starting point would be to define $d(\psi^{(1)},\psi^{(2)})$ such that it provides a measure of the distinguishability between any given pair of states, but in order to distinguish the pair one must make measurements, and one might expect that imposing restrictions on the measurements that are allowed will alter the distinguishability, and thus the distance, between the states.

It is therefore reasonable to expect that there is not a single, universal answer to the question posed at the beginning of the previous paragraph, but rather that the extent to which two states differ from each other will depend on the criteria used \cite{ZyczGeometry}. In this paper, I will consider the statistical distance \cite{Bhatt,WoottersStatDist} between states of a quantum system consisting of subsystems that may be spatially separated from each other, so that any measurement must be restricted to local operations on the individual subsystems along with classical communication between the parties, the LOCC paradigm. We will see that for pure states a restriction to LOCC does not alter the resulting distance between the states. That is, the local and global statistical distances are identical for pure quantum states.

As explained in more detail below, we follow Bhattacharya in defining statistical distance by considering the difference between probability distributions \cite{Bhatt}. Wootters \cite{WoottersStatDist} applied these ideas to the quantum case 
by noting that every outcome of a measurement on a quantum system has a certain probability of occurring, dependent on the initial state of that system, and thus each state becomes associated with a specific probability distribution according to the chosen measurement. By making the same measurement on many copies of the given state, one can determine the probability distribution, and therefore obtain information about the initial state of the system. For the distance between a pair of states, one then seeks an ``optimal" measurement, that which best distinguishes between the states, maximizing the difference between their individual probability distributions. 

In the next section, we describe this approach in more detail, and devise an optimal global measurement that treats the states under consideration in a symmetric way. In Section \ref{atemp}, a brief review of atemporal diagrams \cite{OurAtemp} is given, and then these diagrams are used in Section \ref{local} to prove that a specific local measurement, which is a straightforward generalization of the global one described in Section~\ref{global}, is as good as any global measurement. Finally, we give our conclusions in Section \ref{conclusion}.

\section{Choice of measurement}\label{global}
Our starting point is the Bhattacharya-Wootters distance, $d_c$, on classical probability space,
\begin{equation}
	\label{bhatt-woo}
	\cos\left[d_c(p^{(1)},p^{(2)})\right] = \sum_{i=1}^N \sqrt{p_i^{(1)}}\sqrt{p_i^{(2)}}.
\end{equation}

\noindent For quantum states, the probability distributions will depend on the choice of measurement. For a rank-$1$ positive operator-valued measure (POVM) $\cal M$, with elements corresponding to (possibly non-normalized) states $|\phi_j\rangle$, we have
\begin{equation}
	\label{probs}
	p_j^{(1,2)} = \left|\langle \phi_j|\psi^{(1,2)}\rangle\right|^2.
\end{equation}
The POVM elements satisfy $\sum_j|\phi_j\rangle\langle\phi_j|=I$, with $I$ the identity on the relevant Hilbert space. These considerations lead us to the $\cal M$-dependent distance between the quantum states,
\begin{eqnarray}
	\label{dM}
	\cos\left[d_{\cal M}(\psi^{(1)},\psi^{(2)})\right]  =  \sum_{i=1}^N \left|\langle \phi_i|\psi^{(2)}\rangle\langle \psi^{(1)}|\phi_i\rangle\right| \\
		 \ge  \left|\sum_{i=1}^N \langle \psi^{(1)}|\phi_i\rangle\langle \phi_i|\psi^{(2)}\rangle\right| = \left|\langle \psi^{(1)}|\psi^{(2)}\rangle\right|.\label{bound}
\end{eqnarray}

\noindent The absolute distance between the states, $d(\psi^{(1)},\psi^{(2)})$, for which the above expressions give the upper bound $\cos^{-1}\left[|\langle \psi^{(1)}|\psi^{(2)}\rangle|\right]$, will be given by the most discriminating measurement; or in other words, that measurement which maximizes this distance, minimizing Eq.~(\ref{dM}). It is easily seen that the bound is achieved by choosing a measurement that includes $|\phi_1\rangle = |\psi^{(1)}\rangle$ as one outcome, with $\langle\phi_i|\psi^{(1)}\rangle=0~~\forall i\ne1$. Hence,
\begin{equation}
	\label{probs}
	d(\psi^{(1)},\psi^{(2)}) = \cos^{-1}\left[|\langle \psi^{(1)}|\psi^{(2)}\rangle|\right],
\end{equation}
equal to the angle between the states in Hilbert space. Obviously, another choice of measurement would include $|\psi^{(2)}\rangle$ as one of the outcomes instead of $|\psi^{(1)}\rangle$, and there is in fact a wide range of measurements one could choose from to obtain the bound. This will be important in the local case, since if both these states are entangled it will not be possible to implement a local measurement with any one of the $|\phi_j\rangle$ equal to either of the $|\psi^{(k)}\rangle$. Let us now consider another measurement, one which is optimal in the global case and suggests how to design an optimal local measurement, as well. This measurement, unlike the ones just mentioned, is symmetric with respect to the two states being discriminated.

Consider the operator $|\psi^{(2)}\rangle\langle\psi^{(1)}|$. It is a well-known fact that for this (or any other) operator, there exists an orthonormal basis in which the diagonal elements are all equal to each other \cite{HornJohnson,Walgate}. If we choose the outcomes, $|\phi_i\rangle$, of our measurement to be the (orthonormal) states of the basis that ``equi-diagonalizes" $|\psi^{(2)}\rangle\langle\psi^{(1)}|$, then the quantities $\langle \phi_i|\psi^{(2)}\rangle\langle \psi^{(1)}|\phi_i\rangle$ will be independent of $i$, and the inequality in Eq.~(\ref{bound}) will be satisfied as an equality. Therefore, this measurement is also optimal, achieving the bound.

\section{Atemporal Diagrams}\label{atemp}
The discussion of LOCC measurements may be made relatively transparent by the use of atemporal diagrams \cite{OurAtemp}. In this approach, objects such as bras, kets, and general maps are represented by diagrams. For example,
\begin{equation}
	|\psi_2\rangle =  \includegraphics[scale=1.1,viewport=0 9 5 5]{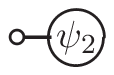}\hspace{.35in}
\end{equation}
denotes a ket, a bra is denoted as $\langle\psi_1| = $  \includegraphics[viewport=0 9 5 5]{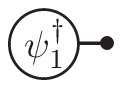} \hspace{.35in}, and an operator (here simply a dyad) as
\begin{equation}
|\psi_2\rangle\langle\psi_1| =  \hspace{0.2in} \includegraphics[viewport=10 9 5 5]{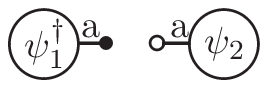}
\end{equation}
\setlength{\baselineskip}{13pt}
\noindent Note that open circles at the end of lines are associated with kets, solid circles with bras. Traces are accomplished by drawing a line connecting an open with a solid circle, which must be associated with the same Hilbert space, such as ${\cal H}_a$ indicated by the label ``a" on the lines in the above diagram; in other words, as is usual, by connecting a bra with a ket to form an inner product. The spatial arrangement and orientation of objects in a given diagram is of no consequence, and I have arranged the above diagram so as to make it easy to connect lines to form traces. That is,
\begin{equation}
\langle\psi_1|\psi_2\rangle =  \hspace{0.2in} \includegraphics[viewport=10 9 5 5]{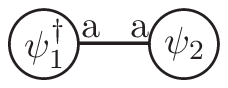}
\end{equation}

\noindent which demonstrates that objects lacking any ``free" ends are simply complex numbers.

To denote states or operators on a multipartite system, we include a line for each subsystem. Our dyad for a bipartite system on ${\cal H}_a \otimes {\cal H}_b$ will then look like

\begin{equation}
\label{bipartite}
\centering
\includegraphics[viewport=100 15 5 5]{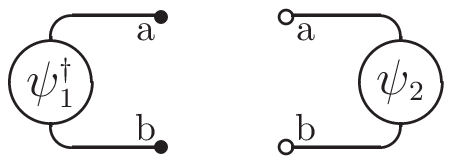}
\end{equation}
\vspace{.025in}

\noindent In the remainder of the paper, the labels on the lines will be omitted with the understanding that the top line corresponds to ${\cal H}_a$, and moving sequentially downward from there we encounter lines for ${\cal H}_b$, ${\cal H}_c$, ${\cal H}_d$, etc., for however many subsystems are involved.

\section{Optimal local protocol}\label{local} 
Now let us turn to our LOCC measurement protocol. The parties will measure sequentially, choosing their measurements based on the outcomes of all previous measurements, which is why the (one-way) classical communication is needed. In a manner similar to the symmetric global measurement described at the end of Section~\ref{global}, each party performs a measurement chosen to equi-diagonalize a particular operator, that operator being the reduction of the dyad $|\psi^{(2)}\rangle\langle \psi^{(1)}|$ to their own subsystem by partially tracing over all the other subsystems that have yet to be measured. For example, suppose in the bipartite case Alice with system ${\cal H}_a$ measures first. She chooses a measurement whose set of outcomes $\{|a_j\rangle\langle a_j|\}$ form the orthonormal basis that brings $\textrm{Tr}_{b}(|\psi^{(2)}\rangle\langle\psi^{(1)}|)$ to a form with all its diagonal elements equal to each other (Tr$_b$ indicates a partial trace over ${\cal H}_b$, only). That is,
\vspace{0.25in}
\begin{eqnarray}\label{bipartite2}
\hspace{-2in}{\large A=A_j=}\hspace{2.25in}
\includegraphics[viewport=150 20 5 5]{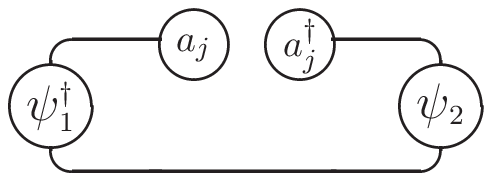}
\end{eqnarray}
\vspace{.05in}

\noindent is a complex number independent of $j$. To see how to obtain this diagram, assuming Alice obtained outcome $j$, then the ket part of $|a_j\rangle\langle a_j|$ attaches to the (${\cal H}_a$) bra part of $|\psi^{(2)}\rangle\langle \psi^{(1)}|$ (line labeled $a$ and having a closed circle in Eq.~(\ref{bipartite})); the bra part of $|a_j\rangle\langle a_j|$ attaches to the ket part of $|\psi^{(2)}\rangle\langle \psi^{(1)}|$ (line labeled $a$ and having an  open circle in Eq.~(\ref{bipartite})); and finally, the trace over ${\cal H}_b$ connects the two lower lines, labeled $b$, of that equation. 

The next step is for Alice to inform Bob that her outcome was $j$, so Bob knows that what remains of the dyad is

\vspace{.15in}
\begin{equation}
\includegraphics[viewport=150 20 5 5]{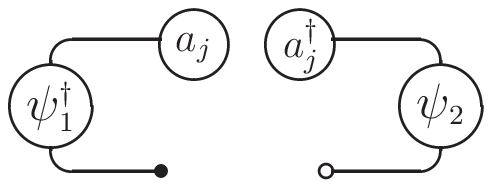}
\end{equation}
\vspace{.05in}

\noindent Therefore, he chooses a $j$-dependent basis $\{|b_k\rangle\}$ ($j$-dependence implicit) to equi-diagonalize this object. If he gets outcome $k$, then together they have

\begin{equation}\label{ajbk}
\hspace{-2in}
\large{B^j=B^j_k=} \hspace{2.25in}\includegraphics[viewport=150 35 5 5]{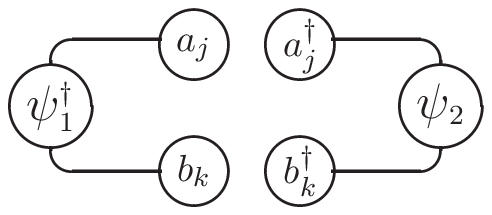}
\end{equation}
\vspace{.15in}

\noindent which is a complex number independent of $k$. I claim that $B^j=B$, independent of $j$, as well. To see this, take the sum of these terms over $k$ to obtain $D_bB^j$, with $D_b$ the dimension of ${\cal H}_b$. This sum is just the partial trace over the space ${\cal H}_b$, which we saw is equal to the quantity $A$ of Eq.~(\ref{bipartite2}). Since $A$ is independent of $j$ because of Alice's choice of measurement, we have that $B^j=B$ is also independent of $j$. 

Now let us consider what has been accomplished. First of all, this last diagram is just $B=\langle \phi_i|\psi^{(2)}\rangle\langle \psi^{(1)}|\phi_i\rangle$, with $i$ a collective index corresponding to $\{j,k\}$ and $|\phi_i\rangle = |a_j\rangle|b_k\rangle$. That is, this diagram represents each of the terms appearing in the sum on the right side of Eq.~(\ref{dM}), but without taking the absolute value. We have seen that with the LOCC measurement described, all of these terms are equal to each other. This means that $\sum_i |B| = |\sum_i B| = |\langle\psi^{(1)}|\psi^{(2)}\rangle|$. Therefore, this LOCC measurement protocol is optimal, achieving the same bound as the best global measurement. 

Now, using the diagrams, we can easily see how this whole business cascades along from one party to the next when there are more than two parties. Imagine that in the previous description, Alice and Bob were just two parties in a long line of parties, so the protocol must continue on. As just argued, we have that the following diagram is a complex number independent of either of their outcomes (this is just Eq.~(\ref{ajbk}) generalized to more than two parties):

\vspace{.15in}
\begin{equation}
\hspace{-2.5in}
\large{B=} \hspace{2.35in}\includegraphics[viewport=165 45 5 5]{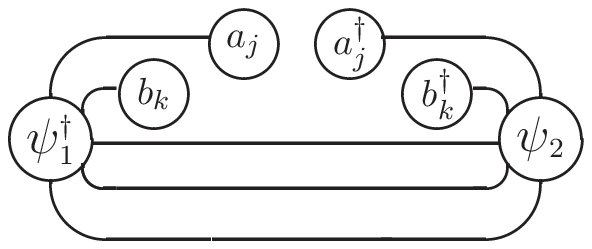}
\label{AB_multi}
\end{equation}
\vspace{.2in}

\noindent  Bob now informs Chloe of both his and Alice's outcomes. Chloe then chooses her measurement basis to equi-diagonalize the dyad shown below,

\vspace{.15in}
\begin{equation}
\includegraphics[viewport=165 45 5 5]{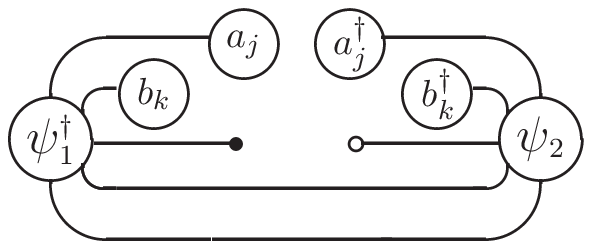}
\end{equation}
\vspace{.175in}

\noindent With outcome $m$, this becomes

\vspace{.175in}
\begin{equation}
\hspace{-2.5in}
\large{C^{jk}_m = C^{jk} = } \hspace{2.35in}\includegraphics[viewport=165 45 5 5]{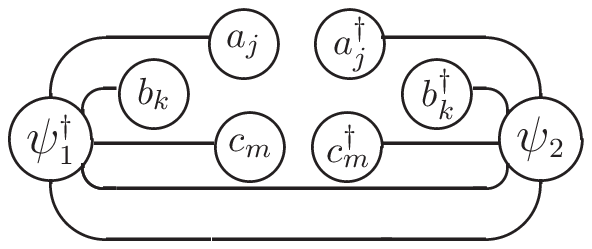}
\end{equation}
\vspace{.25in}

\noindent By Chloe's choice of measurement, this is independent of $m$. Following the same arguments as used above, we have that the sum over $m$ of this quantity is $D_cC^{jk} = B$ (see Eq.~(\ref{AB_multi})), so $C^{jk} = C$, independent of $j$ and $k$. It should be clear that this type of argument will apply to all subsequent measurements regardless of how many parties are involved. Most importantly, this includes the final measurement, which means that the complex number  that is left at the end of the protocol is independent of any of the local measurement outcomes. Since this complex number is equal to the quantity $\langle \phi_i|\psi^{(2)}\rangle\langle \psi^{(1)}|\phi_i\rangle$ appearing on the right-hand side of Eq.~(\ref{dM}), where $|\phi_i\rangle=|a_j\rangle|b_k\rangle|c_m\rangle\dots$, it immediately follows that our LOCC measurement protocol is optimal. That is to say, the best LOCC measurement is equally as good as the best global measurement, and the local statistical distance is identical to the global one.

\section{Conclusion}\label{conclusion}
In summary, we have considered the statistical distance between two pure, multipartite quantum states. We designed a measurement protocol which is symmetric in the sense that it treats the states equally, with outcomes corresponding to projectors onto a basis that equi-diagonalizes the dyad $|\psi^{(1)}\rangle\langle\psi^{(2)}|$, which in the local case also equi-diagonalizes the reductions of this dyad by partial traces over subsets of the parties. These arguments show that the local statistical distance between pure states is identical to the global one.

It is interesting to note that in the case of an orthogonal pair of pure states, the method described above of seeking to equi-diagonalize the dyad $|\psi^{(1)}\rangle\langle\psi^{(2)}|$ brings it to a form where all the diagonal elements vanish, a direct consequence of the fact that this dyad is traceless. This approach would then appear to have a close similarity to that of \cite{Walgate}, who used a measurement basis that ``zero-diagonalizes" a certain operator to show that any pair of orthogonal pure states can be perfectly distinguished by LOCC when only a single copy of the state is available. Indeed, the measurement we have described here is identical to their measurement in this special case of orthogonal pure states. It may be noted that the local measurement of \cite{Walgate} also plays a role in minimizing the probability of error when distinguishing two non-orthogonal pure states in the single-copy scenario \cite{Virmani}. However, the actual optimal measurement used in this case will generally be different than that which is optimal for the statistical distance \cite{FuchsThesis}.

It may also be worth noting that one would not expect the results we've obtained here to hold if quantum mechanics was based on real vector spaces. As argued by Wootters \cite{WoottersReal}, and later by Hardy \cite{HardyReal}, it is generally the case in real-vector-space quantum mechanics that measurements on the parts are not sufficient to determine the state of the whole. It appears, then, that there is something very special about quantum mechanics that makes it possible for local measurements to do as well as global ones. Nonetheless, as discussed in the following paragraph, this conclusion will not always follow even when using the full quantum theory.

Consider the local statistical distance between a pair of mixed states, $d(\rho_1,\rho_2)$. It turns out that there exists a generalization of the method followed in the present paper, where instead of $|\psi^{(2)}\rangle\langle\psi^{(1)}|$, one considers the ``transition operator" $W_1 W_2^\dagger$, with the choice of $W_1$ being such that $\rho_{1}=W_1W_1^\dagger$ and similarly for $W_2$ \cite{Uhlmann1,Uhlmann2,Uhlmann3,AsaBilliards}. However, while the dyad $|\psi^{(2)}\rangle\langle\psi^{(1)}|$ arises naturally from Wootters' analysis for pure states, Eq.~(\ref{bound}), there is no obvious way to see how to obtain $W_1 W_2^\dagger$ from the theory in the mixed state case. In addition, finding a basis that equi-diagonalizes this transition operator will not generally be optimal, at least when there are no restrictions on the measurements that can be used. In this global case, $d(\rho_1,\rho_2)$ has been determined in \cite{FuchsCaves} and shown to be equal to the distance obtained from the Bures-Uhlmann metric \cite{Bures,Uhlmann1}. The arguments of \cite{FuchsCaves} imply that the optimal (global) measurement is unique (except for some special cases \cite{AsaBilliards}), and as this measurement is not generally local, one may conclude that the best local measurement will not do as well. Thus, the local statistical distance will be strictly smaller than the global one when considering mixed quantum states. Finding an optimal local measurement, and thus the local statistical distance between mixed states, remains an interesting open problem.

\section{Acknowledgments}
This work has been supported in part by the National Science Foundation through Grants PHY-0456951 and PHY-0757251, as well as by a grant from the Research Corporation. In addition, I would like to acknowledge the gracious hospitality of the Department of Physics at Lewis and Clark College where this paper was written.

%



\end{document}